\documentclass[12pt,onecolumn,showpacs,preprintnumbers,amsmath,amssymb,%
               preprint]{revtex4}
\usepackage{rotating}
\usepackage{subfigure}
\usepackage{epsfig}
\usepackage{bm}
\usepackage{dcolumn}

\newcommand{\beq}{\begin{equation}}
\newcommand{\eeq}{\end{equation}}
\newcommand{\bea}{\begin{eqnarray}}
\newcommand{\eea}{\end{eqnarray}}

\newcommand{\gcc}{{\rm~g\,cm}^{-3}}

\def\simgr{\,\hbox{\hbox{$ > $}\kern -0.8em \lower 1.0ex\hbox{$\sim$}}\,}
\def\simle{\,\hbox{\hbox{$ < $}\kern -0.8em \lower 1.0ex\hbox{$\sim$}}\,}
\def\beq{\begin{equation}}
\def\eeq{\end{equation}}

\def\simgr{\,\hbox{\hbox{$ > $}\kern -0.8em \lower 1.0ex\hbox{$\sim$}}\,}
\def\simle{\,\hbox{\hbox{$ < $}\kern -0.8em \lower 1.0ex\hbox{$\sim$}}\,}

\def\dd{\mathrm{d}}

\def\araa{ARA\&A~}
\def\apjl{Astrophys.~J.~Letters~}
\def\apjs{Astrophys.~J.~Supp.~}
\def\pr{Phys.\ Rev.~}

\begin{document}

\title{Microcanonical calculations of excess thermodynamic properties of dense binary systems}

\author{Christophe Winisdoerffer}\email{cwinisdo@ens-lyon.fr}
\author{Gilles Chabrier}\email{chabrier@ens-lyon.fr}
\affiliation{\'Ecole normale sup\'erieure de Lyon,
     CRAL (UMR CNRS No.\ 5574),
     69364 Lyon Cedex 07, France}
\author{Gilles Z\'erah}\email{zerah@cea.fr}
\affiliation{Commissariat \`a l'\'Energie Atomique,
BP 12, 91680 Bruy\`eres-le-Ch\^atel,
 France}

\date{\today}


\begin{abstract}

We derive a new formulation to calculate the excess chemical potential of a
fraction of $N_1$ particles interacting with $N_2$ particles of a different species.
The excess chemical potential is calculated numerically from first principles by coupling
molecular dynamics and Thomas-Fermi density functional theory to take
into account the contribution arising from the quantum electrons
on the forces acting on the ions. The choice of this simple functional is motivated by the
fact that the present paper is devoted 
to the derivation and the validation of the method but more complicated functionals can and will be implemented in the future.
This new method is applied in the microcanonical ensemble, the most natural ensemble for
molecular dynamics simulations.
This avoids the introduction of a thermostat in the simulation, and thus uncontroled modifications
of the trajectories calculated from the forces between particles.
The calculations are conducted for
three values of the input thermodynamic quantities, energy and density, and for different total numbers of particles in order to examine the uncertainties due to finite size effects. This method and these calculations lie
the basic foundation to study the thermodynamic stability of dense mixtures, without any {\it a priori} assumption on
the degree of ionization of the different species.

\end{abstract}



\maketitle

\section{Introduction}

The thermodynamic stability of dense ionic mixtures bears important consequences 
not only on our
understanding of the thermodynamic properties of dense binary systems but also 
on the structure
and the evolution of gaseous giant planets. Indeed, the interior of jovian 
planets is composed
essentially of hydrogen and helium under either atomic or 
molecular form in
the outermost envelope and under the form of a partially or fully ionized 
plasma in the inner
regions \cite{guillot95} \cite{hubbard02}. 
Temperatures and pressures along the
Jupiter or Saturn internal isentrope conditions range from about 100 K 
to $\sim$ 20000 K and from
about 1 bar to $\sim 60$ Mbar. Under these conditions, 
not only the hydrogen/helium mixture
experiences pressure ionization but the homogeneous mixture may become 
thermodynamically unstable. 
Such an immiscibility between
helium-rich droplets and a hydrogen-rich fluid will liberate extra
gravitational energy, modifying
significantly the energy balance and thus the cooling of the planet 
\cite{salpeter73} \cite{stevenson77a} 
\cite{stevenson77b}.
For terrestrial applications, inertial confinement fusion experiments or 
laser-driven shock-wave experiments
on hydrogen isotopes reach densities and temperatures characteristic 
of the aforementioned planetary
conditions, probing the thermodynamic properties of dense plasmas and 
requiring a correct theoretical foundation to describe their equilibrium properties.

The theoretical description of the thermodynamic phase diagram of a dense 
two-component system is a particularly complicated task, for it 
requires a correct description of the
{\it excess} free enthalpy (in a pressure-temperature diagram) of the 
mixture with respect to the
pure phases. This excess quantity is very small compared with the 
contributions of both the mixture
and the pure phases (it is by definition close to zero near the critical 
point) and therefore must be
calculated with very high accuracy. Early calculations, based on 
simplified analytic or semi-analytic calculations
of the free energy of the plasma, assumed hydrogen and helium atoms 
to be fully ionized \cite{stevenson75}
\cite{hansen77} \cite{vieillefosse81} \cite{chabrier_guillot}. 
Moreover, these calculations assumed either a rigid electron
background, the so-called binary ionic mixture (BIM) model, or a 
polarizable electron background within
the linear response approximation. Although correct at very 
high density or temperature, these
assumptions fail when electrons and protons start to recombine.
The phase diagrams calculated
under these conditions are thus restricted to a
reduced (high) density-temperature range. Further attempts 
to do a 
consistent, first-principle determination
of the H/He phase diagram, with no assumption on the electron 
distribution around the ionic centers and a correct
treatment of the various $N$-body ion and electron interactions, 
were based either on extrapolation at
finite-temperature of zero-temperature calculations \cite{klepeis91} 
or on incorrect thermodynamics integration \cite{pfaffen95} 
and thus remain also of doubtful validity. 
Under such circumstances, it is clear that not only
the thermodynamic phase diagram of a hydrogen-helium system at high 
density has not been established
accurately yet, but the correct calculation of the excess free enthalpy of a 
concentration of atoms immersed in an interacting system of different species remains to be done.

In this paper, we derive a new method to address this very point, which is crucial for a 
reliable determination of the thermodynamic phase
diagram  of dense binary mixtures. This method is applied in the microcanonical ensemble and  allows the
direct calculation, from first principles, of the excess chemical potential of 
a binary mixture of nuclei and electrons interacting through the Coulomb potential.
Calculations in the microcanonical ensemble
allow a
fully consistent calculation between the forces acting on the particles and 
the induced trajectories, without the introduction of thermostats. 
We first derive the 
thermodynamic equations which allow the
exact determination of the excess chemical potential. We then combine the 
density functional theory (DFT) to describe the
quantum mechanical properties of the electrons and molecular dynamics (MD) 
to integrate the ion classical equations
of motion
to calculate this chemical potential. Since the present paper is devoted 
to the derivation of the method, we use a simplified functional form
for the electrons, namely the Thomas-Fermi approximation, in order to 
speed up the
minimization of the energy. We limit our calculations to three different values 
of the appropriate input thermodynamic
quantities, namely energy and density in the microcanonical formulation 
used in the present paper. In a future work, devoted to the global 
analysis of the H/He mixture under
various thermodynamic conditions, a more general functional form will be 
implemented. 
Section 2 presents the derivation of the
chemical potential of $N_1$ atoms of a given species
interacting with $N_2$ nuclei of a different species. Section 3 describes our
general energy functional
to take into account the quantum
behaviour of the electrons when computing the ionic configurations, a necessary
condition for an accurate treatment of the problem.
Section 4 is
devoted to the description of the MD numerical computations,
to the discussion of the finite size effects and to the presentation
of the results obtained for different thermodynamic conditions.
The last section is devoted to the conclusion.

\section{Derivation of the excess chemical potential}

As mentioned in the introduction, the ultimate goal of our calculations is
to determine the thermodynamic stability of a given number $N_1$ of atoms immersed in a system of $N_2$
particles of a different species under given thermodynamic conditions, without any assumption on 
the electron distribution around the nuclei, {\it i.e.} on the degree of ionization of the atoms.
The stability of such a mixture involves the calculation of the mixing 
enthalpy of the system, {\it i.e.} of
the excess chemical potential of each immersed atom.
The chemical potential $\mu_i$ of a particle $i$ immersed in a plasma
corresponds by definition to the change of the state function of
the appropriate thermodynamic ensemble when one adds or removes this particle to/from the plasma.
When the thermodynamic limit is achieved $(N\rightarrow \infty, V
\rightarrow \infty$, $N/V=$constant$)$, the result does not depend
either on the ensemble or on the fact that the particle has been added to
or removed from the surrounding plasma. Because of the large 
fluctuations of the system away from its equilibrium configuration,
one can not add or remove directly a particle, in particular at
high density or if the interaction potential is too stiff.
The correct approach consists in modifying
progressively the interaction potential $\lambda \,V(r)$ between the particle
$i$ and the
surrounding particles $j\ne i$. The case $\lambda=0$ corresponds 
to the case where the particle $i$
does not interact with the other 
particles, but retains its discernability character (case of an ideal mixture), whereas the case 
$\lambda=1$ corresponds to the sought two-component system with full 
interactions. This method illustrates the so-called thermodynamic integration 
approach. In the
microcanonical ensemble, with fixed energy, volume and number of 
particles $(E,{\cal V},N)$, 
the chemical potential of a particle ``1" of mass $m_1$ corresponds to the
calculation of the following expression derived
in appendix A-C:

\begin{eqnarray}
-\frac{\mu_1}{kT}&=&\ln \left[ \left( \frac{2\pi m_1}{h^2} \right)^{3/2}
\left( \frac{\sum_{j=1}^{N}m_j}{\sum_{j=1}^{N+1}m_j} \right)^{3/2}
\frac{1}{N_1+1} \frac{\Gamma\left(\frac{3(N-1)}{2}\right)}{\Gamma\left(
\frac{3N}{2}\right)}
{\cal V} \langle K_N^{\frac{3}{2}} \rangle \right] \nonumber \\
&+&\left( \frac{3N}{2}-1 \right)
\int_{\lambda=0}^{\lambda=1} {\mathrm{d}}\lambda \left \langle \frac{1}{E(\lambda)-V}
\frac{\partial E(\lambda)}{\partial \lambda} \right \rangle
,
\label{eq_chem_pot}
\end{eqnarray}

\noindent where $\Gamma$ is the Gamma function, ${\cal V}$ is the cell volume,
$K_N$ is the kinetic energy and $E(\lambda)$ is
the energy corresponding to the system with the interaction potential 
$\lambda \,V(r)$.
The brackets $\langle ... \rangle$ denote a microcanonical average.
The first term on the right hand side is the ideal part of the 
chemical potential, arising from 
the entropy cost due to the
particle insertion or removal, while the second term represents the non-ideal 
contribution of the chemical potential,
which depends on the interaction between the
particle under consideration and the rest of the system.
The integral can be estimated by a Gauss-Legendre quadrature \cite{numrec}:

\begin{eqnarray}
\int_{\lambda=0}^{\lambda=1} {\mathrm{d}} \lambda \left \langle 
\frac{1}{E(\lambda)-V}
\frac{\partial E(\lambda)}{\partial \lambda} \right \rangle
\simeq \frac{1}{2}\sum_{i=1}^{n} \omega_i \left\langle \frac{1}{E(\lambda)-V}
\frac{\partial E(\lambda)}{\partial \lambda} \right\rangle _{\lambda_i=
\frac{x_i+1}{2}}
,
\label{eq_chem_pot_quad}
\end{eqnarray}

\noindent where $x_i$ are the zeros of Legendre polynomials and $\omega_i$ 
are the associated weights \cite{abramovitz}.

These calculations, in practice,
require great caution. By definition
of the microcanonical ensemble, the total (potential+kinetic) energy of 
the system must be conserved
along the simulation. As a consequence, at each
step where the potential goes from $\lambda_i V$ to $\lambda_{i+1} V$,
the kinetic part of the energy must be renormalized in order to
maintain the total energy constant.
Therefore, the correct calculation of the chemical potential consists in generating
several particle configurations (to be described in \S 4) corresponding to the Hamiltonian $H+\lambda V$, and
in computing the averages which appear in Eq.(\ref{eq_chem_pot}). This
can be done by a fully classical simulation if
the interaction between each kind of particles is described by a classical 
2-body potential between particles. Such an approach, however, can not take into
account the fact that the potentials strongly depend on the density and
the temperature, evolving from a potential characteristic of atoms at low density and temperature
to a long-range Coulomb potential characteristic of a fully ionized plasma at high
density and/or temperature. Therefore, the correct phase diagram, without
any assumption on the interaction potentials, requires
{\it ab initio} generations of representative ionic configurations. This approach is presented in
the next sections.

\section{The functional of the electrons}

As mentioned in the introduction, a correct study of the problem under 
consideration
requires a correct treatment of the ion and electron interactions. This implies to take into account
the effects of the quantum nature of the electrons on the forces 
acting on the ions.
Since the pioneering work of Hohenberg and Kohn \cite{hk64}, many problems
involving interacting electrons have been tackled within the framework of
the density functional theory (DFT). This theory turns the
diagonalization problem of a many-body Hamiltonian into the minimization
of a functional $\Omega[n({\bm r})]$ of the electron density, 
a much easier approach when dealing with many electrons.
For this reason, the DFT has been extensively used
in condensed matter and is described in detail in many textbooks 
(see e.g.~\cite{parr}). 
In the framework of the DFT, the grand potential  of the
electrons can be written in the form: 

\begin{eqnarray}
\Omega[n({\bm r})]=\int {\mathrm{d}} {\bm r}\, (V_{\mathrm{ie}}({\bm r})-\mu) \, n({\bm r}) +
F[n({\bm r})]
,
\end{eqnarray}

\noindent where $F[n({\bm r})]$ is a universal functional of the ground state density
$n({\bm r})$ of the interacting electrons and 
$V_{\mathrm{ie}}({\bm r})$ denotes the external ion-electron potential.
$\Omega [n({\bm r})]$ is minimum when $n({\bm r})$ corresponds to the correct  density.
In our calculations, we have chosen to write $\Omega[n({\bm r})]$ under
the following simplified form (in order to speed up the minimization):

\begin{eqnarray}
\Omega[n({\bm r})] & = & k_B T \int {\mathrm{d}} {\bm r} n({\bm r})
\frac{F_{3/2}(\eta)}{F_{1/2}(\eta)} +
c_{\mathrm{ex}} \int {\mathrm{d}} {\bm r} [n({\bm r})]^{4/3}+
E_{\mathrm{corr}}[n({\bm r})] \nonumber \\
& + & c_W \int {\mathrm{d}} {\bm r} \frac{[\nabla n({\bm r})]^2}
{n({\bm r})}+\frac{e^2}{2}\int \!\! \int {\mathrm d}{\bm r_1}
{\mathrm d}{\bm r_2}\frac{n({\bm r_1})n({\bm r_2})}
{|{\bm r_1}-{\bm r_2}|}  \nonumber \\
& + & \int {\mathrm{d}} {\bm r} n({\bm r})
\left(V_{\mathrm{\mathrm{ie}}}({\bm r}, {\bm R_{\mathrm{ion}}})-\mu \right)
,
\label{eq_tffunc}
\end{eqnarray}

\noindent 
where $c_{\mathrm{ex}}=-\frac{3}{4}(3/\pi)^{1/3}
e^2$ is the exchange Dirac coefficient, $c_W=(\sigma/8)\hbar^2/m_e$
is the von Weizs\"acker gradient correction coefficient 
\cite{parr} \cite{weiz35} (with $\sigma=1$ in our case), and
$E_{\mathrm{corr}}[n({\bm r})]$ is given by a parametrization \cite{perdew92}
of Monte Carlo simulations \cite{ceperley80}.
$F_{3/2}(\eta)$ and $F_{1/2}(\eta)$ are the Fermi integrals, where
$\eta=(\mu-V({\bm r}))/k_BT$  is obtained by the inversion of the relation:
$n({\bm r})=(2\pi^2)^{-1}(2m_e/\hbar^2)^{3/2}(k_BT)^{3/2}F_{1/2}(\eta)$.
Accurate fitting formulae of the Fermi integrals and inverse integrals have been
published in the literature \cite{blinnikov96} \cite{antia93}.
For a given configuration of the nuclei, we are able to find the
electronic density $n({\bm r})$ which corresponds to the ground state of the system.
The Hellmann-Feynman theorem \cite{feynman39} \cite{nielsen85} enables us to calculate the forces arising 
from this electron density acting
on the nuclei and the stress tensor on a cell (of which diagonal terms 
correspond to the pressure components).  \\
Within the Born-Oppenheimer approximation, we can thus make a classical molecular dynamics (MD)
simulation of the nuclei sub-system, while taking into account in the calculation
of the forces the quantum behaviour of the electrons.
The Born-Oppenheimer approximation expresses the fact that the 
electrons respond
instantaneously to a change of configuration of the ions, a fairly good 
assumption for dense ionic systems. The calculation of 
the last term on the right hand side
of Eq.(\ref{eq_tffunc}), {\it i.e.} the interaction with the external ionic 
potential, involving effective
pseudopotentials, is described below.

\section{Molecular dynamics}

\subsection{Method}

We have computed Eq.(\ref{eq_chem_pot}) for a number of $N_1$ helium
atoms of nuclear charge $Z_1=2$ and mass $M_1$ immersed in a system of $N_2$ hydrogen particles of charge $Z_2=1$ and mass $M_2$. The thermodynamic averages
in Eq.(\ref{eq_chem_pot}) are estimated by generating a set of representative
configurations of the system in a cubic reference cell of size $L$ with periodic boundary conditions. This is done by a dynamical simulation of the equations of motion for the ions:

\begin{eqnarray}
M_i \frac{{\mathrm{d}}^2{\bm R_i}}{{\mathrm{d}}t^2}={\bm F_i}
,
\label{eqn_motion}
\end{eqnarray}

\noindent where $M_i$ is the mass of the $i$th nuclei.
The forces ${\bm F_i}$ between particles (or equivalently the total potential $V({\bm r})$)
arising from electron and ion $N$-body interactions, beyond any linear 
approximation for the electron-induced
screening effects of the core potential,
are calculated from a density functional approach. These forces involve the ones arising
from the quantum electron distribution obtained from Eq.(\ref{eq_tffunc}) and
the ones derived from the interionic potential $Z_iZ_je^2/{|{\bm R_i}-{\bm R_j}|}$.
The equations of motion are solved with a standard Verlet velocity 
algorithm \cite{frenkel}.
The crucial point of the present paper is that these calculations are completed
in the microcanonical ensemble, {\it i.e.} at constant energy, volume and total
momentum \cite{note1}. 
Standard simulations, in other thermodynamic ensembles, imply the introduction
of a thermostat, either by reinitializing the velocities ``periodically'' or by introducing new degrees of freedom. These thermostats, however, yield a perturbation of the trajectories, which no longer represent the ones determined by the forces. Such unphysical effects are avoided in the present
microcanonical calculations, which insure full consistency between the forces and the trajectories.

The {\it ab initio} calculations, with the aforedescribed functional, have been performed with the 
ABINIT code \cite{abinit}.
We replace the bare Coulomb potential of the nucleus by a pseudopotential, which differs from the true Coulomb 
potential below a cutoff radius $r_{{\mathrm{loc}}}$, removing the cusp constraint at $r\rightarrow 0$
and avoiding the $1/r$ singularity.
The pseudopotentials used in our
simulations are those of Hartwigsen {\it et al.} for helium \cite{hgh98},
and Goedecker {\it et al.} for hydrogen \cite{gth96}.
These pseudopotentials are constructed so as to reproduce with high accuracy
the Kohn-Sham free energy.

The aforementioned cutoff radius determines an upper bound in density for the domain of validity of the pseudopotentials.
The ones used in the present calculations~\cite{note2} 
have $r_{{\mathrm{loc}}}=0.2$ bohr,
which implies a density limit: $a\simgr r_{{\mathrm{loc}}}$,
{\it i.e.} $r_s \simgr 0.2$, where $r_s=a/a_0$ is the density parameter, 
$a=(\frac{V}{4\pi N/3})^{1/3}$ 
is the mean distance between nuclei
and $a_0$ is the Bohr radius. This
condition corresponds to
$\rho\simle 335\, \gcc$.

\begin{figure}[h]
\centerline{\includegraphics[height=7.2cm]{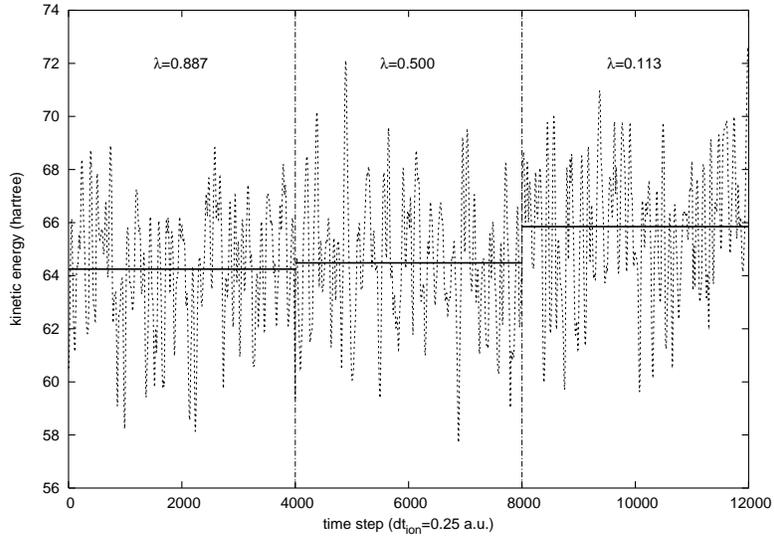}}
\caption{Kinetic energy during the whole switch off of the helium atom
embedded in a 63 hydrogen system, in the 3-point quadrature case.
The average kinetic energies are displayed in solid line, the instantaneous
ones into dashed line. The vertical lines separate the different
domains of constant switching parameter $\lambda$.}
\label{fig_ekin}
\end{figure}

As mentioned earlier, in order to maintain the total energy constant during
the process of switching on or off the interaction, the kinetic contribution $E_{\mathrm{kin}}$
must be renormalized. For the
thermodynamic conditions of our runs, this corresponds to a decrease of 
$E_{\mathrm{kin}}$ because the potential energy increases as the interaction
is switched on
($\lambda =0\rightarrow 1)$. This implies a large initial kinetic energy. 
The condition is more easily fulfilled if one chooses to switch {\it off} 
the interaction instead of switching it on ($\lambda =1\rightarrow 0)$.
Indeed, during such a process, the kinetic part must be increased (instead 
of decreased), which is
always possible. 
As a consequence of this renormalization, we can not associate an accurate
temperature until the thermodynamic limit is reached. 
This process is represented on Figure \ref{fig_ekin}, which displays the 
kinetic energy during 
the whole process, and shows the discontinuities
appearing in the mean value of $E_{\mathrm{kin}}$ when $\lambda$ changes from 
$\lambda_i$ to $\lambda_{i+1}$.

\subsection{Results}

\begin{figure}[h]
\centerline{\includegraphics[height=7.2cm]{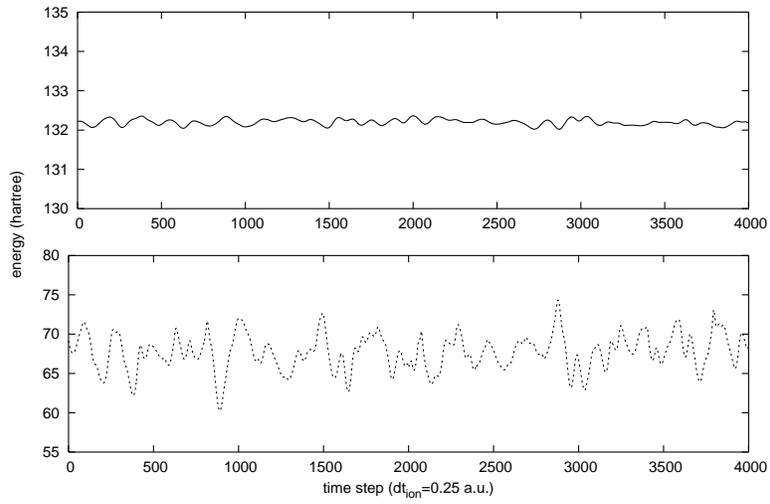}}
\caption{Total (solid line) and potential (dotted line) energies corresponding
to the simulation of $\{$63 H, 1 He$\}$ with the switching parameter
equal to 0.5.}
\label{fig_e22}
\end{figure}

We have tested our procedure on a system consisting of 63 hydrogen nuclei
and 1 helium nucleus.
The thermodynamic conditions of our microcanonical simulation are: 
$E_{\mathrm{tot}}=
132.21$ hartree and $V=L^3=57.906 $ bohr$^3$, which 
correspond to $r_s\simeq 0.6$,
$T\simeq 2\; 10^5$ K and $P\simeq 7\; 10^4$ GPa.
The reference cell of the simulation assumes
periodic boundary conditions, with one particle exiting the cell on one 
side replaced by one entering the
opposite side. 
In order for the final results not to depend on the initial distribution,
which corresponds to a random
distribution of the positions and velocities (obtained by a classical
molecular dynamics simulation to prevent atoms to overlap),
we let the system relax during 4000 time steps, a very conservative limit for the considered
densities and temperatures.
Even though the main contribution to the total energy at $r_s=0.6$ comes from
the nearly uniform electron background, the forces depend partly on
the non-uniformity of this electron density distribution. 
Therefore, in order to calculate the forces correctly, the electronic
density, more precisely the departure of the density from a homogeneous
distribution, must be calculated with very high accuracy. In order to
fulfil this condition, we require the energy to converge within 
$|\Delta E/E|< 10^{-8}$.
Unfortunately, high accuracy in the functional minimization does not
preclude energy fluctuations during the simulation due to the
discretization of the Newton equations and, most importantly, to
finite-size effects. These points are examined below.
The equations of motion are solved using
the Verlet algorithm with a time step equal to $dt_{ion}=0.25$ a.u.
This time step enables us to resolve the dynamics
of our system accurately for any value $\lambda \, V$ of the He-H interaction.
Figure \ref{fig_e22}
displays the conservation of the total energy obtained in 
our simulation
with this time step.
The integral Eq.(\ref{eq_chem_pot_quad})
is first calculated
with a $M$-point Gauss-Legendre quadrature.
\noindent
After the first 4000 time steps to let the system relax, an
other 4000 time steps simulation is ran to generate several configurations.
The argument $\partial_{\lambda}E(\lambda)/(E(\lambda)-V)$ is calculated numerically, by
calculating $E(\lambda\pm 0.01)$ every 10 time steps for a fixed
configuration and $E(\lambda)-V$. A last run is devoted to the calculations of the ideal
part of the chemical potential by generating 10000 different configurations
of 63 H
and the evaluation of $\langle K_N^{3/2}\rangle$. 

\begin{figure}[h]
\centerline{\includegraphics[height=8cm]{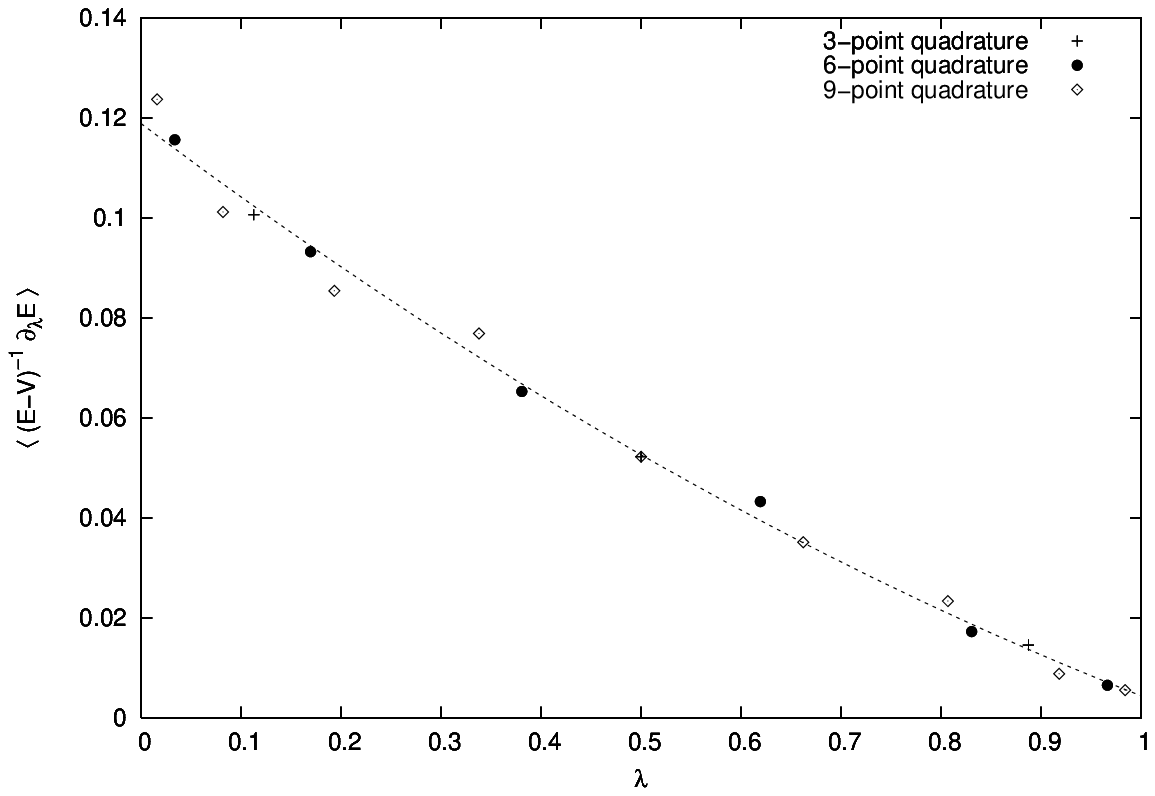}}
\caption{Different values of $\left\langle \frac{1}{E(\lambda)-V}
\frac{\partial E(\lambda)}{\partial \lambda} \right\rangle$
obtained for the different quadratures. A quadratic fit of the
results is given as a guide for the eye.}
\label{fig_quadra}
\end{figure}

We have tested the validity of the $M$-point quadrature to estimate the
integral Eq.(\ref{eq_chem_pot_quad}) by doing similar calculations, for the same
thermodynamic conditions, with a 3-point, 6-point and a 
9-point quadrature. The results are shown on Figure \ref{fig_quadra}, and
the resulting evaluations of the integral are given in Table \ref{tab-Gauss}.
As seen in this table, a 6-point quadrature is enough to 
calculate accurately the integral (\ref{eq_chem_pot_quad}). Our calculations of
the chemical potential $-\frac{\mu_1}{kT}$
of a helium atom embedded in a 63-H plasma
for our thermodynamic conditions ($-\frac{\mu_1^0}{kT}$ corresponds
to the ideal part of the chemical potential and $-\frac{\mu_1^{1}}{kT}$ to the 
excess contribution) yield:

\begin{eqnarray}
\left\{
\begin{array}{l}
\vspace*{0.2cm}
\displaystyle -\frac{\mu_1^0}{kT}=\displaystyle 14.02 ,\\
\vspace*{0.2cm}
\displaystyle -\frac{\mu_1^{1}}{kT}=\displaystyle 5.14 ,\\
\vspace*{0.2cm}
\displaystyle -\frac{\mu_1}{kT}=\displaystyle 19.16 .
\end{array}
\right.
\end{eqnarray}

In order to estimate the $N$-dependence of our result, we have also
calculated the entropy cost which corresponds to the removal of 2 He-particles surrounded by
126 H, and 4 He-particles surrounded by 252 H, for
the same thermodynamic conditions (density and energy) as for
the \{1 He, 63 H\} system.
These computations are much more
time consuming (the computation time scales roughly as $t\propto N^3$),
and the removal of the helium
atom must be done 2 and 4 times, respectively. We expect
a very small dependence of the results on the size ($N$) of the simulated system. Indeed,
we do not calculate the chemical potential of {\it one} He atom
but the chemical potential of a constant $\it fraction$ 
($x_{\mathrm{He}}=$1/64) of He in a H-He mixture.
As a consequence, the $N$-dependence of the results stems from
the interaction between a He atom with its replicas (due to
the periodic condition boundaries)
and from the fact that the
accessible phase space increases with the total number of atoms.
The first effect becomes important when
the characteristic length of interaction is of the same order as
the simulated box, {\it i.e.} at much higher density than the present simulations.
Quantification of the second effect requires
simulations with different values of $N$.
The results are presented in Table \ref{tab-Ndep}. The chemical potential is estimated
with a centered scheme, {\it i.e.} the chemical potential corresponding
to a Helium fraction $x_{\mathrm{He}}=$1/128=0.008 is evaluated from the entropy difference between
the \{0 He, 63 H\} system and the \{1 He, 63 H\} system (with full
interaction). The statistical uncertainties
on  $-\mu_1/kT$ are $\pm 0.02$ for
the \{1 He, 63 H\} and \{2 He, 126 H\} systems, and $\pm 0.04$ for
the \{4 He, 252 H\} one (achieving the same statistical uncertainties
scales as $N^3$). 
The results between the three systems
for a He fraction equal to 0.008, as shown in the Table, are thus fully compatible, and 
no statistically-significant trend appears. The same simulations yield also
the estimation of the Helium chemical potential for different
number fractions. All the results are given in Table \ref{tab-Ndep}, and are compatible
within the statistical uncertainties.
For the \{1 He, 63 H\} mixture, we have also conducted calculations for two other thermodynamic conditions, displayed in Table \ref{tab-diffthermo}. As expected
intuitively,
it is easier to add an atom in a low density plasma than in a high density one
(at constant total energy), or in
a cold plasma than in a hot one (at constant density).

It is interesting to compare our results with the limit of rigid
electronic background at high density for the binary ionic mixture,
the so-called BIM limit
\cite{chabrier98} \cite{potekhin00}. Our reference conditions,
$E_{\mathrm{tot}}= 132.21$ hartree and $V=L^3=57.906 $ bohr$^3$, {\it i.e.}
$r_s\simeq 0.6$, correspond to
$T= 2.2\; 10^5$ K, $P= 6.7\; 10^4$ GPa
and
$\Delta S=19.15\,k_B$ (Table \ref{tab-Ndep}) in the simulation. For this density and
temperature, the BIM corresponds to $P=7.2\; 10^4$ GPa and
$\Delta S =21\,k_B$.
We have also ran a simulation at higher density, namely $r_s=0.3$, which is
close to the density limit of our pseudopotentials. The total energy is equal to
$E_{\mathrm{tot}}= 672.76$ hartree.
In that case, the present calculations yield $T=4.5\; 10^5$ K, 
$P= 2.15\; 10^6$ GPa,
whereas the BIM results are $T=4.5\; 10^5$ K, $P= 2.2\; 10^6$ GPa.
The small differences between the simulations and the BIM
reflect the contribution due to the electron gas polarization
(inhomogeneous distribution),
which starts playing a role around these densities, and the contribution due
to the interactions between particles of {\it different species} (namely H and He). These latter
are not taken into account in the BIM, which is based
on the so-called linear volume law, where only the ideal entropy of
mixture is included.


\begin{table}
\caption{Integration of Equation \ref{eq_chem_pot_quad}.}
\label{tab-Gauss}
\begin{ruledtabular}
\begin{tabular}{ccccc}
\  & 3 $\,$ points & 6$\,$ points & 9 $\,$ points & trapeze \\
\noalign{\smallskip}
\hline
\noalign{\smallskip}
$\displaystyle \int_{\lambda=0}^{\lambda=1} {\mathrm{d}}\lambda \left \langle
\frac{1}{E(\lambda)-V} \frac{\partial E(\lambda)}{\partial \lambda}
\right \rangle$ & 0.06568 & 0.05578 & 0.05548 & 0.05565  \\
\end{tabular}
\end{ruledtabular}
\end{table}


\begin{table}
\caption{Finite size effects on the chemical potential.}
\label{tab-Ndep}
\begin{ruledtabular}
\begin{tabular}{cccccccccc}
System  & \multicolumn{3}{c}{\{1 He, 63 H\}} & 
     \multicolumn{3}{c}{\{2 He, 126 H\}} & 
     \multicolumn{3}{c}{\{4 He, 252 H\}} \\
\noalign{\smallskip}
\cline{2-4}  \cline{5-7} \cline{8-10}
\noalign{\smallskip}
$x_{\mathrm{He}}=N_{\mathrm{He}}/(N_{\mathrm{H}}+N_{\mathrm{He}})$ & 0.004 & 0.008 & 0.012 & 0.004 & 0.008 & 0.012 &
                         0.004 & 0.008 & 0.012 \\
\noalign{\smallskip}
\hline
\noalign{\smallskip}
$\displaystyle -\frac{\mu_1}{kT}$ & x     & 19.16 & x &
                                    19.15 & 19.15 & 19.15 &
				    19.17 & 19.14 & 19.21 \\
\end{tabular}
\end{ruledtabular}
\end{table}


\begin{table}
\caption{Chemical potentials for different thermodynamic conditions.}
\label{tab-diffthermo}
\begin{ruledtabular}
\begin{tabular}{cccc}
\{$E_{\mathrm{tot}}$(hartree), $V$(bohr$^3$)\} & \{132.21,57.91\} & \{132.21,139.40\} &
\{20.06,139.40\} \\
$r_s$ & 0.6 & 0.8 & 0.8 \\
\noalign{\smallskip}
\hline
\noalign{\smallskip}
$\displaystyle -\frac{\mu_1}{kT}$ & 19.16 & 16.59 & 13.94 \\
\end{tabular}
\end{ruledtabular}
\end{table}

\section{Conclusion}

In this paper, we have derived a new method, based on physics first principles, to calculate the
excess potential of a number fraction of particles immersed in a mixture of particles of different species,
as given by Eq.(\ref{eq_chem_pot}).
The calculations are based on a consistent
treatment of the forces acting on the nuclei, taking into account the contribution arising from
the quantum electrons, by calculating self-consistently the equations of motion of the classical
nuclei and the functional density of the electronic distribution. The method is applied directly
in the microcanonical ensemble, avoiding the use of a thermostat, and thus insures consistency between the forces and the trajectories of the particles. The bare Coulomb potential is approximated at
short distances by pseudopotentials which remain valid up to large densities ($r_s \simgr 0.2$), where the linear response theory becomes valid. The thermodynamic quantities are calculated for different configurations, representing the evolution
of the interaction, and thus of the system,
from the initial case of an {\it ideal} atom ``1" inserted in a system of particles ``2" to the 
final case where all interactions between the immersed particle and the surrounding
nuclei are taken into account. Only properly following such a series of changes of equilibrium states  insures thermodynamic consistency and thus allows a correct evaluation of the energy and
pressure contributions to the excess chemical potential of the immersed particle.
Previous simulations \cite{pfaffen95} calculated the excess enthalpy directly from the difference beween the final and the initial states, yielding an incorrect evaluation of the contraction work, and thus
of the pressure contribution.

The validity of the method has been tested with the case of a dense hydrogen/helium mixture
for three different helium fractions and three different thermodynamic states.
The forces are calculated with very high accuracy,
with a convergence criterium $|\Delta E/E|< 10^{-8}$.
Finite size effects
on the final results have been quantified
and found to be small ($\sim 10^{-3}$)
leading to fluctuations of the same order on the total energy (see Fig 2). 
The method provides 
robust foundations for 
accurate evaluations of the excess
thermodynamic quantities of dense binary mixtures, without any assumption on the electron density distribution and thus on the degree of ionization of the atoms. This opens the door to accurate calculations of phase diagrams of dense
mixtures of atoms and partially or fully ionized plasmas, a subject of prime interest for the structure and the
evolution of giant gaseous planets. Work in this direction is in progress.

\vspace*{0.4cm}
We are very grateful to G\'erard Massacrier and Alexander Potekhin for very useful discussions and insightful remarks. We are also indebted to the two anonymous referees for their valuable comments which helped improving the initial manuscript. 

\appendix
\begin{widetext}

\section{Microcanonical average}

We consider a (classical) system with fixed total energy $E$,
volume $V$ and total momentum ${\bm p_{\mathrm{tot}}}$, which contains two different kinds of
particles, $N_1$, $N_2$, with $N=N_1+N_2$.
In this microcanonical ensemble, the number of accessible states
for this system is:

\begin{eqnarray}
\delta \Omega = \frac{\delta E}{h^{3N}N_1 ! N_2 !} \int \dd {\bm p}^N \dd {\bm q}^N \; \delta (E-H) \; \delta ({\bm p_{\mathrm{tot}}}-\sum_{j=1}^N {\bm p_j})
,
\end{eqnarray}
where $H=\sum_{j=1}^N \frac{{\bm p_j}^2}{2m_j}+V({\bm q}^N)$ is the Hamiltonian
of the ionic centers, and includes the modification of the Coulomb potentials due to the electron gas polarization, $d{\bm p}^N = \prod_{i=1}^{N} \prod_{j=1}^{3} dp_{ij}$, and $dp_{ij}$
is the j-component of the momentum of the particle i. Idem for $d{\bm q}^N$.
Then:
\begin{eqnarray}
\Omega = \frac{1}{h^{3N}N_1 ! N_2 !} \int \dd {\bm p}^N \dd {\bm q}^N \; \theta(E-H) \; \delta ({\bm p_{\mathrm{tot}}}-\sum_{j=1}^N {\bm p_j}) \equiv \frac{1}{h^{3N}N_1 ! N_2 !} \, I(E,{\bm p_{\mathrm{tot}}})
,
\end{eqnarray}
\end{widetext}
where $\theta(x)=1$  if $x>0$, 0 otherwise, and $I(E,{\bm p_{\mathrm{tot}}})$ denotes the integral. \\

The Laplace transform (toward E) of I is \cite{note3}:

\begin{eqnarray*}
{\cal L}[I] & = &  \int \dd {\bm p}^N \dd {\bm q}^N \; \delta ({\bm p_{\mathrm{tot}}}-\sum_{j=1}^N {\bm p_j}) \int_{\mathrm{min}(H)}^{+\infty} \dd E \; \theta(E-H) e^{-sE} \\ & = & \int \dd {\bm p}^N \dd {\bm q}^N \; \delta ({\bm p_{\mathrm{tot}}}-\sum_{j=1}^N {\bm p_j}) \frac{1}{s} e^{-sH} \\
& = & \int \dd {\bm p}^N \; \delta ({\bm p_{\mathrm{tot}}}-\sum_{j=1}^N {\bm p_j}) e^{-s\sum_{j=1}^N \frac{{\bm p_j}^2}{2m_j}} \; \; \int \dd {\bm q}^N \; \frac{1}{s} e^{-sV({\bm q}^N)}
.
\end{eqnarray*}

The Hamiltonian $H$ is general and does not have, in particular, to be positive, although it
needs to have a lower limit.
Note that  with a change of variable $\zeta=E-\mathrm{min} (H)$, the integral $\int_{\mathrm{min}(H)}^{+\infty}$ becomes $\int_0^{+\infty}$, and the results remain unchanged.

\noindent
Then:
\begin{eqnarray*}
J_0 & = & \int \dd {\bm p}^N \; \delta ({\bm p_{\mathrm{tot}}}-\sum_{j=1}^N {\bm p_j}) e^{-s\sum_{j=1}^N \frac{{\bm p_j}^2}{2m_j}}  \equiv  J^3
,
\end{eqnarray*}
where $J=\int \dd p_x^N \delta (p_{tot_x}-\sum_{j=1}^N p_{j_x}) e^{-s\sum_{j=1}^N \frac{p_{j_x}^2}{2m_j}}$. 

\vspace*{0.2cm}
\noindent
$J$ can be calculated by Fourier transformation:
\begin{eqnarray*}
{\cal F}[J] & = & \int \dd p_x^N \; \frac{1}{\sqrt{2\pi}} \int_{-\infty}^{+\infty} \dd p_{tot_x} \delta (p_{tot_x}-\sum_{j=1}^N p_{j_x}) \; \exp \left(-s\sum_{j=1}^N \frac{p_{j_x}^2}{2m_j}\right) \; \exp(i\zeta p_{tot_x}) \\
& = & \frac{1}{\sqrt{2\pi}} \prod_{j=1}^{N} \int \dd p_{j_x} \exp\left(-s \frac{p_{j_x}^2}{2m_j}+i\zeta p_{j_x}\right) \\
& = & \frac{1}{\sqrt{2\pi}} \left( \prod_{j=1}^{N} \sqrt{\frac{2\pi m_j}{s}} \right) \exp\left(-\frac{\sum_{j=1}^N m_j}{2s} \zeta ^2\right)
.
\end{eqnarray*}

\noindent
Then:
\begin{eqnarray*}
J & = & {\cal F}^{-1}[{\cal F}[J]] =
\frac{\prod_{j=1}^{N} \sqrt{2\pi m_j}}{\sqrt{2\pi \sum_{j=1}^N m_j}} \frac{1}{s^{(N-1)/2}} \exp \left(-\frac{s}{2\sum_{j=1}^N m_j} p_{tot_x}^2\right)
.
\end{eqnarray*}

\noindent
With ${\bm p_{\mathrm{tot}}}=0$, we get:
\begin{eqnarray*}
J_0=J^3= \left( \frac{\prod_{j=1}^{N} \sqrt{2\pi m_j}}{\sqrt{2\pi \sum_{j=1}^N m_j}} \right) ^3 \frac{1}{s^{\frac{3(N-1)}{2}}}
,
\end{eqnarray*}
and:
\begin{eqnarray*}
I={\cal L}^{-1}[{\cal L}[I]]=\left( \frac{\prod_{j=1}^{N} \sqrt{2\pi m_j}}{\sqrt{2\pi \sum_{j=1}^N m_j}} \right) ^3 \int \dd {\bm q}^N \; \theta(E-V({\bm q}^N))\frac{(E-V({\bm q}^N))^{\frac{3(N-1)}{2}}}{\Gamma\left(\frac{3(N-1)}{2}+1\right)}
.
\end{eqnarray*}

\noindent
Equation (A2) thus reads:
\begin{eqnarray}
\Omega(E,{\bm p_{\mathrm{tot}}}=0) & = & \frac{1}{h^{3N}N_1 ! N_2 !} \left( \frac{\prod_{j=1}^{N} \sqrt{2\pi m_j}}{\sqrt{2\pi \sum_{j=1}^N m_j}} \right) ^3 \int \dd {\bm q}^N \; \theta(E-V({\bm q}^N))\frac{(E-V({\bm q}^N))^{\frac{3(N-1)}{2}}}{\Gamma\left(\frac{3(N-1)}{2}+1\right)}
. \nonumber \\
& &
\end{eqnarray}

\noindent
In a similar way, we have for $\omega \widehat{=} \left( \frac{\partial \Omega}{\partial E} \right)_{V,N_1,N_2}$:
\begin{eqnarray}
\omega =  \frac{1}{h^{3N}N_1 ! N_2 !} \left( \frac{\prod_{j=1}^{N} \sqrt{2\pi m_j}}{\sqrt{2\pi \sum_{j=1}^N m_j}} \right) ^3 \int \dd {\bm q}^N \; \theta(E-V({\bm q}^N))\frac{(E-V({\bm q}^N))^{\frac{3(N-1)}{2}-1}}{\Gamma\left(\frac{3(N-1)}{2}\right)}
.
\end{eqnarray}

\noindent
For a quantity depending only on ${\bm q}^N$, {\it i.e.} $A({\bm q}^N)$, we 
have:
\begin{eqnarray}
\nonumber \langle A \rangle & \equiv & \frac{1}{\omega} \frac{1}{h^{3N}N_1 ! N_2 !} \int \dd {\bm p}^N \dd {\bm q}^N \; \delta (E-H) \; \delta ({\bm p_{\mathrm{tot}}}-\sum_{j=1}^N {\bm p_j}) \; A({\bm q}^N) \\
& = & \frac{\int \dd {\bm q}^N \; \theta(E-V({\bm q}^N))(E-V({\bm q}^N))^{\frac{3(N-1)}{2}-1} A({\bm q}^N)}{\int \dd {\bm q}^N \; \theta(E-V({\bm q}^N))(E-V({\bm q}^N))^{\frac{3(N-1)}{2}-1}}
,
\end{eqnarray}
where $\langle ... \rangle$ denotes a {\it microcanonical average}.

\section{Chemical potential by the particle insertion method}

The definition of the chemical potential of the particle 1 is:
\begin{eqnarray}
-\frac{\mu_1}{kT} & = & \left( \frac{\partial S}{\partial N_1} \right)_{E,V,N_2}
.
\end{eqnarray}

\noindent
With $S=k\ln \omega$ and the equations derived in Appendix A, this yields:
\begin{eqnarray}
-\frac{\mu_1}{kT} & = & \frac{\ln \omega_{N_1+1}-\ln \omega_{N_1}}{N_1+1-N_1} = \ln \frac{\omega_{N_1+1}}{\omega_{N_1}} \nonumber \\
& = & \ln \left[ \left( \frac{2\pi m_1}{h^2} \right)^{3/2} \left( \frac{\sum_{j=1}^{N}m_j}{\sum_{j=1}^{N+1}m_j} \right)^{3/2} \frac{1}{N_1+1} \frac{\Gamma\left(\frac{3(N-1)}{2}\right)}{\Gamma\left(\frac{3N}{2}\right)} \frac{\int \dd {\bm q}^{N+1} \; \theta(E-V)(E-V({\bm q}^{N+1}))^{\frac{3N}{2}-1}}{\int \dd {\bm q}^N \; \theta(E-V)(E-V({\bm q}^N))^{\frac{3(N-1)}{2}-1}}\right]
.\nonumber \\
& &
\end{eqnarray}

\noindent
Let define $I_N$ as:
\begin{eqnarray*}
I_N & = & \int \dd {\bm q}^N \; \theta(E-V({\bm q}^N)) (E-V({\bm q}^N))^{\frac{3(N-1)}{2}-1} \\
& \equiv & \int \dd {\bm q}^N \; \theta(E-V({\bm q}^N)) K_N^{\frac{3(N-1)}{2}-1}
,
\end{eqnarray*}
where $K_N \equiv E-V({\bm q}^N)$ is a function of ${\bm q}^N$ and should not 
be formally confused with the kinetic part of the Hamiltonian (even if $K_N$
is equal to the kinetic energy in a molecular dynamics simulation).

\noindent
We get:
\begin{eqnarray*}
I_{N+1} & = & \int \dd {\bm q}_{N+1} \; \int \dd {\bm q}^N \; \theta(E-V) (E-V({\bm q}^N)-V({\bm q}_{N+1}))^{\frac{3N}{2}-1} \\
& = & \int \dd {\bm q}_{N+1} \; \int \dd {\bm q}^N \; \theta(E-V) K_N^{\frac{3}{2}} K_N^{\frac{3(N-1)}{2}-1} \left( 1-\frac{V({\bm q}_{N+1})}{K_N} \right) ^{\frac{3N}{2}-1}
,
\end{eqnarray*}
and:
\begin{eqnarray*}
\frac{I_{N+1}}{I_N} = \int \dd {\bm q}_{N+1} \left\langle K_N^{\frac{3}{2}} \left( 1-\frac{V({\bm q}_{N+1})}{K_N} \right)^{\frac{3N}{2}-1} \right\rangle
,
\end{eqnarray*}

\noindent
which yields for the chemical potential:
\begin{eqnarray}
\nonumber -\frac{\mu_1}{kT} & = & \ln \left[ \left( \frac{2\pi m_1}{h^2} \right)^{3/2} \left( \frac{\sum_{j=1}^{N}m_j}{\sum_{j=1}^{N+1}m_j} \right)^{3/2} \frac{1}{N_1+1} \frac{\Gamma\left(\frac{3(N-1)}{2}\right)}{\Gamma\left(\frac{3N}{2}\right)} \int \dd {\bm q}_{N+1} \left\langle K_N^{\frac{3}{2}} \left( 1-\frac{V({\bm q}_{N+1})}{K_N} \right)^{\frac{3N}{2}-1} \right\rangle \right] .\\
& &
\end{eqnarray}

\section{Chemical potential by the thermodynamic integration method}

The variation of entropy when going from a state with interaction $\lambda =0$ to
$\lambda=1$ reads:
\begin{eqnarray}
\Delta S = \int_{\lambda=0}^{\lambda=1} \dd \lambda \; \frac{\partial S}{\partial \lambda}
.
\end{eqnarray}

\noindent
We can thus derive:
\begin{eqnarray*}
\frac{1}{k} \frac{\partial S}{\partial \lambda} & = & \frac{1}{\omega} \frac{\partial \omega}{\partial \lambda} \\
& = & \frac{\partial_{\lambda}\int \dd {\bm q}^{N+1} \; \theta (E(\lambda)-V({\bm q}^{N+1}))(E(\lambda)-V({\bm q}^{N+1}))^{\frac{3N}{2}-1}}{\int \dd {\bm q}^{N+1} \; \theta (E(\lambda)-V({\bm q}^{N+1}))(E(\lambda)-V({\bm q}^{N+1}))^{\frac{3N}{2}-1}} \\
& = & \left( \frac{3N}{2}-1 \right) \frac{\int \dd {\bm q}^{N+1} \; \theta (E(\lambda)-V({\bm q}^{N+1}))\frac{1}{E-V}\frac{\partial E}{\partial \lambda}(E(\lambda)-V({\bm q}^{N+1}))^{\frac{3N}{2}-1}}{\int \dd {\bm q}^{N+1} \; \theta (E(\lambda)-V({\bm q}^{N+1}))(E(\lambda)-V({\bm q}^{N+1}))^{\frac{3N}{2}-1}} \\
& = & \left( \frac{3N}{2}-1 \right) \left\langle \frac{1}{E-V} \frac{\partial E}{\partial \lambda} \right\rangle
.
\end{eqnarray*}

\noindent
The thermodynamic integration proceeds in two steps. The first one deals 
with the insertion of a free particle into the system. The entropy cost
of this insertion is given by the formula established for the insertion
method. 
\begin{eqnarray}
-\frac{\mu_1^0}{kT} & = & \ln \left[ \left( \frac{2\pi m_1}{h^2} \right)^{3/2} \left( \frac{\sum_{j=1}^{N}m_j}{\sum_{j=1}^{N+1}m_j} \right)^{3/2} \frac{1}{N_1+1} \frac{\Gamma\left(\frac{3(N-1)}{2}\right)}{\Gamma\left(\frac{3N}{2}\right)} {\cal V} \langle K_N^{\frac{3}{2}} \rangle \right],
\end{eqnarray}
where ${\cal V}$ is the cell volume.

\noindent
The interaction is then progressively switched on, and the non-ideal
part of chemical potential is then given by:
\begin{eqnarray}
-\frac{\mu_1^1}{kT} & = & \left( \frac{3N}{2}-1 \right) \int_{\lambda=0}^{\lambda=1} \dd \lambda \left\langle \frac{1}{E(\lambda)-V} \frac{\partial E(\lambda)}{\partial \lambda} \right\rangle
.
\end{eqnarray}

\noindent
The total chemical potential is the sum of the two contributions:
\begin{eqnarray}
-\frac{\mu_1}{kT} = -\frac{\mu_1^0}{kT} -\frac{\mu_1^1}{kT}
.
\end{eqnarray}



\newpage
 
\newpage
  
\newpage


\begin{thebibliography}{}
\bibitem{note1} In order to correct for numerical errors in the molecular dynamics simulation, it is convenient to fix also momentum, since the total momentum of the box must be conserved. In the thermodynamic limit, both ensembles lead to the same results.
\bibitem{note2} Even if atomic units are the most natural ones in the context
of DFT, cgs ones may be more convenient in a different context; here are some
conversion factors: 1 hartree $\simeq 4.36 \; 10^{-11}$ erg, 1 bohr
$\simeq 0.53 \; 10^{-8}$ cm, and 1 atomic unit of time
$\simeq 2.42 \; 10^{-17}$ s.
\bibitem{note3} The function $I(x)$ is continuous by steps and
is bound by $\exp (ax)$ for $x\rightarrow +\infty$ so that the Lapace transform can be applied to $I(x)$.
\end{thebibliography}
\end{document}